\documentstyle [12pt]{article}
\begin{document}
\author{{\it L.B.Litinsky}}
\title{{\bf Highly Symmetric Neural Networks 
of Hopfield Type (exact results)}}
\date{{\small Institute for High Pressure Physics Russian Academy of Sciences,
142092 Troitsk, Moscow region, Russian Federation,  
e-mail: litin@ns.hppi.troitsk.ru}}
\maketitle
\begin{abstract}
A set of fixed points of the Hopfield type neural network is under
investigation. Its connection matrix is constructed with regard to the Hebb
rule from a highly symmetric set of the memorized patterns.
Depending on the external parameter the analytic description of the fixed
points set has been obtained. 
\end{abstract}
A set of fixed points of the Hopfield type neural network is under
investigation. Its connection matrix is constructed with regard to the Hebb
rule from a $(p\times n)$-matrix $\bf S$ of memorized patterns:
$${\bf S}=\left(\begin{array}{ccccccc}
1-x&1&\ldots&1&1&\ldots&1\\
1&1-x&\ldots&1&1&\ldots&1\\
\vdots&\vdots&\ddots&\vdots&\vdots&\ldots&\vdots\\
1&1&\ldots&1-x&1&\ldots&1\end{array}\right).$$
Here $n$ is the number of neurons, $p$ is the number of memorized
patterns $\vec s^{(l)}$, which are the rows of the matrix $\bf S$, and $x$ is
an arbitrary real number.
 
Depending on $x$ the memorized patterns $\vec s^{(l)}$ are interpreted as 
$p$ distorted vectors of the {\it standard}
$$\vec\varepsilon (n)= (\underbrace{1,1,\ldots,1}_n).\eqno(1)$$

The problem is as follows: {\it the network has to be learned by $p$-times
showing of the standard (1), but a distortion has slipped in the learning 
process.
How does the fixed points set depends on the value of this distortion $x$?}

Depending on the distortion parameter $x$ the analytic description of the 
fixed points set has been obtained. It turns out to be very important that the
memorized patterns $\vec s^{(l)}$ form a highly symmetric group of vectors: 
all of them correlate one with another in the same way:
$$
(\vec s^{(l)},\vec s^{(l')})=r(x),  \eqno (2)
$$
where $r(x)$ is independent of $l,l'=1,2,\ldots,p.$ Namely this was the reason 
to use the words "highly symmetric" in the title.

It is known [1], that the fixed points of a network of our kind have to be of
the form:
$$\vec\sigma^*=(\sigma_1,\sigma_2,\ldots,\sigma_p,1,\ldots,1),\quad
\sigma_i=\{\pm 1\},\ i=1,2,\ldots,p.\eqno(3)$$
Let's join into one {\it class} $\Sigma^{(k)}$ all the {\it configuration}
vectors $\vec\sigma^*$ given by Eq.(3), which 
have $k$ coordinates equal to "--1"
among the first $p$ coordinates. The class $\Sigma^{(k)}$ consists of 
$C_p^k$ configuration vectors of the form (3), and there are $p+1$ 
different classes $(k=0,1,\ldots,p)$. Our main result can be formulated as a 
Theorem.

{\bf Theorem.} {\it As $x$ varies from $-\infty$ to $\infty$ the fixed
points set is exhausted in consecutive order by the classes of the vectors 
$$\Sigma^{(0)},\Sigma^{(1)},\ldots,\Sigma^{(K)},$$
and the transformation of the fixed points set from the class $\Sigma^{(k-1)}$
into the class $\Sigma^{(k)}$ occurs when $x=x_k$:
$$x_k= p\frac{n-(2k-1)}{n+p-2(2k-1)},\quad k=1,2,\ldots,K.$$
If $\frac{p-1}{n-1}<\frac13$, according this scheme all the $p$ 
transformations
of the fixed points set are realized one after another and $K=p$. If  
$\frac{p-1}{n-1}>\frac13 $, the transformation related to
$$K=\left[\frac{n+p+2}4\right]$$
is the last. The network has no other fixed points} 

The Theorem makes it possible to solve a number of practical problems. 
We would
like to add that the Theorem can be generalized onto the case 
of arbitrary vector 
$$\vec u =(u_1,u_2,\ldots,u_p,1,\ldots,1),\quad \sum_{i=1}^p u_i^2=p$$ 
being a standard instead the standard (1). Here memorized patterns 
$\vec s^{(l)}$ are
obtained by the distortion of the first $p$ coordinates of the vector $\vec u$
with regard to the fulfillment of Eqs.(2).

The obtained results can be interpreted in terms of neural networks,
Ising model and factor analysis.
\vskip 2mm
[1] L.B.Litinsky. Direct calculation of the stable points of a 
neural network. Theor. and Math. Phys.{\bf 101}, 1492 (1994)
\end{document}